\documentclass{article}
\usepackage{ijcai24}

\usepackage{times}
\usepackage{soul}
\usepackage{url}
\usepackage[hidelinks]{hyperref}
\usepackage[utf8]{inputenc}
\usepackage[small]{caption}
\usepackage{graphicx}
\usepackage{amsmath}
\usepackage{amsthm}
\usepackage{booktabs}
\usepackage{algorithm}
\usepackage{algorithmic}
\usepackage[switch]{lineno}
\usepackage{newtxmath} 
\usepackage{multirow, makecell}
\usepackage{dsfont}
\usepackage{subfigure}
\usepackage{enumerate}

\newtheorem{theorem}{Theorem}
\newtheorem{definition}{Definition}
\newtheorem{lemma}{Lemma}
\newtheorem{remark}{Remark}
\newcommand{\nash}{MPNE-BBS}
\newcommand{\buy}{b}

\newcommand{\Sel}{\mathscr{S}}
\newcommand{\Buy}{\mathscr{D}}

\DeclareMathOperator*{\argmax}{arg\,max}
\DeclareMathOperator*{\argmin}{arg\,min}

\graphicspath{ {./figures/} }

\title{Optimizing Prosumer Policies in Periodic Double Auctions Inspired by Equilibrium Analysis \\ \textsc{Extended version}\footnote{This is an extended version of paper accpeted at IJCAI 2024} }

\author{
Bharat Manvi$^1$
\and
Sanjay Chandlekar$^{1,2}$\and
Easwar Subramanian$^1$\\
\affiliations
$^1$TCS Research, $^2$IIIT Hyderabad\\
\emails
\{bharat.manvi,  easwar.subramanian\}@tcs.com,
sanjay.chandlekar@research.iiit.ac.in}

\begin{document}

\maketitle

\begin{abstract}
We consider a periodic double auction (PDA)  wherein the main participants are wholesale suppliers and brokers representing retailers. The suppliers are represented by a composite supply curve and the brokers are represented by individual bids. Additionally, the brokers can participate in small-scale selling by placing individual asks; hence, they act as prosumers \footnote{Prosumers are entities that can buy and sell.}. Specifically, in a PDA,  the prosumers who are net buyers have multiple opportunities to buy or sell multiple units of a commodity with the aim of minimizing the cost of buying across multiple rounds of the PDA. Formulating optimal bidding strategies for such a PDA setting involves planning across current and future rounds while considering the bidding strategies of other agents. In this work, we propose Markov perfect Nash equilibrium (MPNE) policies for a setup where multiple prosumers with knowledge of the composite supply curve compete to procure commodities. Thereafter, the MPNE policies are used to develop an algorithm called {\nash } for the case wherein the prosumers need to re-construct an approximate composite supply curve using past auction information. The efficacy of the proposed algorithm is demonstrated on the PowerTAC wholesale market simulator against several baselines and state-of-the-art bidding policies.
\end{abstract}

\section{Introduction}
\label{sec:intro}

   Auctions play a crucial role in the real world, serving as dynamic marketplaces wherein the value of a commodity gets determined by supply and demand in real-time ~\cite{auction_wiki}. Double auctions are the most prominent type of auction, with a trade volume of more than a trillion dollars daily in stock exchanges~\cite{Parsons2011} and energy markets~\cite{Ketter2020}. Such auctions involve bids from the buyers and asks from the sellers; these bids and asks are submitted as a tuple comprising the desired unit price for the commodity and the intended number of units for procurement (or sale). Subsequently, the auction mechanism determines each participant's clearing price and quantity. 
    
   A periodic double auction (PDA) is a setup wherein buyers and sellers engage in a (finite) sequence of auctions to trade certain units of a commodity. For example, a buyer with the intention of procuring certain units of a commodity will engage in multiple (but finite) trades with the seller to satisfy her desired procurement target. These types of auctions are prevalent in energy markets wherein a power generating company (GenCo) that sells energy in bulk is the prominent seller and retail brokers are the buyers. In addition, PDAs are also used to model the call auctions in financial markets \cite{9680552}.  The PDA setup allows buyers and sellers to trade energy periodically until a few hours before delivery. As players need to participate in a series of auctions,  devising a bidding strategy that caters to current as well as future auctions is a challenging problem. Considering that energy trades in a smart grid setup enable trades worth $1000$ $TWh$ of energy, resulting in daily transactions of more than $3$ billion dollars just in Europe \cite{nordpool}, it is prudent to design efficient bidding strategies on behalf of a market participant to bring in cost optimization and ecosystem efficiency. To this end, we model PDAs as a finite horizon Markov game and propose equilibrium solutions to aid in devising efficient bidding strategies. 

   Developing efficient bidding strategies depends mainly on clearing and payment rules of the auction mechanism. Although equilibrium solutions for double auctions have been studied under various payment rules \cite{Wilson1992Chapter8S} including some existence results for average clearing price rule \cite{Satterthwaite1989}, multi-buyer \cite{susobhan20} and multi-item settings \cite{sanjay22}, all these works involve simple double auctions and hence may not be applicable for PDAs as devising bidding strategies in PDA involves sequential decision-making. Some recent works do model PDAs as a Markov game and use techniques from reinforcement learning such as multi-agent Q learning \cite{Rashedi2016}, multi-agent deep Q network \cite{Ghasemi2020}, deep deterministic policy gradients \cite{Du2021,sanjay22}. Nevertheless, many of these works involve single-sided auctions\footnote{Auctions wherein only buy or sell bids are placed} and involve numerical simulations rather than analytical solutions. A recent work ~\cite{manvi2023nash} did propose an equilibrium solution by modelling the PDA as a complete information Markov game. However, the work is limited to cases where the clearing mechanism is the average clearing price rule (ACPR) and supply by wholesale suppliers is adequate. Furthermore, the players involved in the auction were only buyers, not the prosumers. Moreover, the said work does not provide any policy for a realistic auction setting, which involves incomplete information.
   
   Our work overcomes the abovementioned limitations by providing equilibrium strategies for prosumers (not just buyers) for any general uniform clearing rule satisfying certain properties. We then propose a novel bidding strategy {\nash} inspired by this analytical equilibrium solution that approximately reconstructs the supply curve from past auction information to place bids in the market. Our distinct contributions are as follows,

   \begin{enumerate}
        \item We consider a Markov game framework wherein the retail market brokers (net buyers or prosumers) can also participate in small-scale selling, enabling retailers with solar panels and EVs to sell energy along with buying. 
       \item We define the properties of the uniform clearing rule of the double auction, which encompasses various clearing mechanisms prevailing in diverse auction setups across different geographies.
       \item We propose novel Markov perfect Nash equilibrium (MPNE) solutions when prosumers compete to procure the required commodities. The proposed solutions are valid for any uniform clearing mechanism of the PDA with the above mentioned properties.
       \item Based on the solution concept derived for the complete information setting, we propose a novel bidding strategy, \nash, to work in the \textit{incomplete information} setup where the supply curve and information regarding the demand requirement of other buyers are not known. 
       \item We then demonstrate the efficacy of the \nash\ algorithm against several baseline and state-of-the-art bidding strategies deployed in the close-to-real-world energy market simulator PowerTAC. 
   \end{enumerate}  

\section{Related Work}
\label{sec:rel_work}

    Some of the early work in the double auction setting was devoted to obtaining equilibrium solutions for single buyer and single seller with a uniform distribution of valuations~\cite{chatterjee1983}. \cite{Satterthwaite1989} attempt to find non-trivial equilibria and show the existence of a multiplicity of equilibria for the $k$-double auction for a generic class of market participants' valuations. They propose the equilibrium strategies in the form of differential equations and then examine the efficiency of the proven equilibrium. A work \cite{krausz1997markov} considers analytical solutions for a two-player zero-sum Markov game of incomplete information; whereas this work considers the multi-player general-sum game. \cite{vetsikas2014} proposes equilibrium strategies for multi-unit sealed bid auction for $m^{th}$ and $(m+1)^{th}$ price sealed bid auction, which differ from $k$-double auctions in clearing rules.  

    PowerTAC \cite{Ketter2020} is a widely adopted platform to validate bidding strategies in PDAs and has a vast literature on bidding strategies. Various works have proposed Markov Decision Process (MDP) based strategies; for instance, \cite{tactex2013}'s MDP-based strategy was inspired by Tesauro and Bredin's bidding strategy \cite{tesauro2002}, which they solve using dynamic programming. Urieli and Stone's strategy was improved upon by 
 ~\cite{susobhan20}, where the authors also provide equilibrium analysis for single-item single-shot double auctions. ~\cite{sanjay22} present an analytical equilibrium solution for single-shot multi-unit auctions to design a DDPG-based bidding strategy. \cite{rodrigue2013} also proposed an MDP-based bidding strategy to determine the bid quantity and use of Non-Homogeneous Hidden Markov Models (NHHMM) to determine bid prices. Additionally, some works have also adopted a Monte Carlo Tree Search (MCTS) framework to device bidding strategies for PDAs~\cite{spot18,tuctac2021}.
    
At best, all the strategies mentioned above involve equilibrium analysis of single-shot double auctions. Hence, the analysis may not be readily extended to the PDA setting which is a multi-shot auction and the current work bridges this gap. 

\section{Market Clearing Mechanism}\label{sec:MCM}

    We begin by describing the market clearing mechanism of a double auction. Consider a group of $N$ prosumers who want to procure multiple units of a commodity from a group of sellers by participating in a PDA having $H$ rounds. At any round $h \in [H]$\footnote{For any integer $K$, we denote $[K]$ as the set $\{1,\dots,K\}$.} of the PDA, a prosumer $\buy \in [N]$ has  $Q^{\buy,h}_{+}$  and $Q^{\buy,h}_{-}$ units of a commodity to buy and sell, respectively. We let $Q^{\buy,h}_{+}, Q^{\buy,h}_{-}$ to be unique $\forall b \in [N]$ with  $Q^{\buy,h}_{+}  > Q^{\buy,h}_{-} $. The set of buy bids is denoted as $\mathcal{B}^{\buy,h}_{+}$ with $B^{\buy,h}_{+}$ elements and the sell bids\footnote{Sell bids are prosumer's asks.} are denoted by $\mathcal{B}^{\buy,h}_{-}$ with $B^{\buy,h}_{-}$ elements. The buy bids are the pair of price and quantity denoted as $(p^{b,h}_{i,+},q^{b,h}_{i,+}), \; i \in [B^{\buy,h}_{+}]$, where price and quantity are bounded by $p_{\max}$ and $Q^{\buy,h}_{+}$ respectively. Similarly, the sell bids are denoted as $(p^{b,h}_{n,-},q^{b,h}_{n,-}), \; n \in [B^{\buy,h}_{-}]$, with $p_{\max}$ and $Q^{\buy,h}_{-}$ as price and quantity upper bounds. As a result, the total outstanding demand in round $h \in [H]$ from all $N$ prosumers is denoted as $Q^{\Buy,h} = \sum\nolimits_{\buy\in[N]} Q^{\buy,h}_{+}$.

    The wholesale sellers, on the other hand, are represented by a consolidated supply curve with $L^h$ asks expressed as  $\mathcal{L}^{h} = \{(p^{h}_i,q^{h}_i) \; | \; i \in [L^h]\}$, with $p^h_i \in [0,p_{\max}]$ and $q^h_i \in [0,q_{\max}]$ as the price and quantity components of the $i^{\rm th}$ ask $(p^h_i,q^h_i)$, with $p_{\max}$ and $q_{\max}$ as suitable upper bounds. Accordingly, the total supply provided by wholesale sellers, at round $h$, is denoted as $Q^{\Sel,h} = \sum\nolimits_{m \in [L^h] }q^h_m $. Now, the overall supply at round $h$ of the PDA is given by  $\hat{Q}^{\Sel,h} = Q^{\Sel,h}+ Q^{\Sel,h}_{-}$, where $Q^{\Sel,h}_{-} = \sum\nolimits_{\buy \in [N]} Q^{\buy,h}_{-}$ is the supply from the prosumers. Finally,  at the round $h$, the combined asks and sell bids of the wholesale suppliers and the brokers is expressed as $\hat{\mathcal{L}}^h = \{(\hat{p}^{h}_i,\hat{q}^{h}_i) \; | \; i \in [L^h + B^h_{-}]\}$. 

    The market regulator at each round $h$, collects the elements of $\hat{\mathcal{L}}^h$ and  $\mathcal{B}^h_{+}$ and uses a \textit{clearing rule} to produce a clearing price(s) and cleared quantities. In this work, we focus on the \emph{uniform clearing rule}, where all the cleared bids and asks have the same clearing price. 
    The cleared price for round $h$ is called the market clearing price (denoted as $\lambda^h$) and the total quantity cleared for all the $B^h_{+}$ bids at round $h$ is denoted as $Q^h$. To further elaborate the clearing process, we assume, without loss of generality, that the elements of the set $\hat{\mathcal{L}}^{h}$ ($\mathcal{B}^{h}_{+}$) are sorted in increasing (decreasing) order of the price component. If the price components of the elements are equal, then $\hat{\mathcal{L}}^h$ and $\mathcal{B}^{h}_{+}$ are sorted in decreasing order of the quantity component. If both price and quantity components of certain elements are equal, then the ordering between them is arbitrary. More concretely, we consider the uniform clearing rules, which have properties defined in Definition \ref{def:CM}. 

\begin{definition} \label{def:CM}
        Given 
        \begin{itemize}
            \item $(\hat{p}^{h}_{d},\hat{q}^{h}_{d}) \in \hat{\mathcal{L}}^h$ as the last cleared ask and $(p^{\buy,h}_{l,+},q^{\buy,h}_{l,+}) \in \mathcal{B}^h_{+}$ as the last cleared buy bid at round $h$.
            \item $\alpha^{\buy,h}_i$ with $ i  \in [B^{\buy,h}_{+}]$ and $\beta^{\buy,h}_k$ with $k \in [B^{\buy,h}_{-}]$ as the cleared buy and sell bid quantities of a prosumer  $\buy \in [N]$ at round $h$.
        \end{itemize}
        The \textbf{properties} of the considered uniform clearing rules are defined as follows.
        \begin{itemize}   
            \item  \label{prop:clearing_price} The uniform clearing price $\lambda^h$ at round $h$ is a scalar value that lies in the interval $ [\hat{p}^{h}_d,p^{\buy,h}_{l,+}]$.  
            \item \label{prop:total_cl_qty}  The total market cleared quantity at round $h$ is,
            \begin{align*}
                Q^h = \min\left\{\sum\nolimits_{j=1}^{d} \hat{q}^{h}_j, \sum\nolimits_{i=1}^{l} q^{\buy,h}_{i,+} \right\}.
            \end{align*}
             \item Buy bids that are higher than the last cleared buy bid are fully cleared. That is , $\alpha^{\buy,h}_m = q^{\buy,h}_{m,+}$, if $p^{\buy,h}_{m,+} > p^{\buy,h}_{l,+}$.
             \item Buy bids that are lower than the last cleared buy bid are not cleared. That is, $\alpha^{\buy,h}_m = 0$,  if  $p^{\buy,h}_{m,+} < p^{\buy,h}_{l,+}$.
             \item Buy bids that are \textit{equal} to the last cleared buy bid are cleared as 
            $\alpha^{\buy,h}_{m} = \frac{1}{|B_{b=l}|} \left(Q^h - \sum\nolimits_{j=1}^{|B_{b>l}|}q^{\buy,h}_{j,+} \right),$ where $|B_{b=l}|$ and $|B_{b>l}|$ denote the number of bids that are equal and higher than the last cleared bid, respectively. 
            \item Conversely, the asks and sell bids from $\hat{\mathcal{L}}^h$ are cleared such that the cheaper asks are given the higher priority. 
        \end{itemize}     
    \begin{remark}\label{rem:bid_sell_no_match}
        In the clearing mechanism, we assume that the buy bid of player $\buy$ cannot be matched with her own sell bid when all the elements of $\hat{\mathcal{L}}^h$ are sell bids of player $\buy$. 
       \end{remark}
    \end{definition}
    In practice, regulators deploy a variety of clearing rules that satisfy the properties defined in Definition \ref{def:CM}. Popular examples include \textit{merit order dispatch} \cite{taylor2015convex} and $k$-double auctions \cite{Angaphiwatchawal2021}. The merit order dispatch is a mechanism where the market operator clears the market by maximizing the area between supply (asks) and demand curves (bids). Specifically, market clearing is posed as an optimization problem, and the (primal and dual) solutions to this optimization give the cleared quantities and clearing price. The clearing price for  the $k$-double auction (where $k \in [0\;;1]$) is given by $\lambda^h = k \cdot \hat{p}^{h}_d + (k-1) \cdot p^{\buy,h}_{l,+}$. Furthermore, the average clearing price mechanism (ACPR) is a special case of $k$-Double auction with $k=0.5$, and its clearing price is given as the average of the last cleared ask and bid. Further details of the clearing mechanism are as follows. 

\begin{figure}[h]
    \centering
    \subfigure[]{\includegraphics[width=0.45\linewidth]{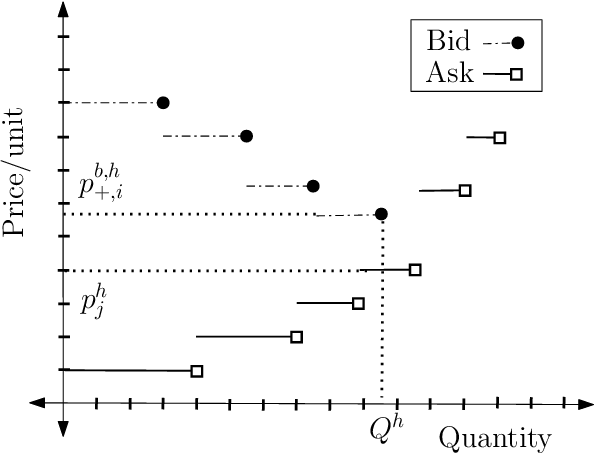}}
    \subfigure[]{\includegraphics[width=0.45\linewidth]{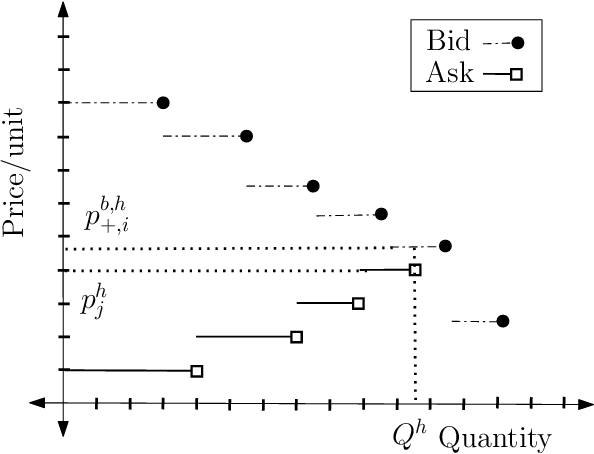}}
    \subfigure[]{\includegraphics[width=0.45\linewidth]{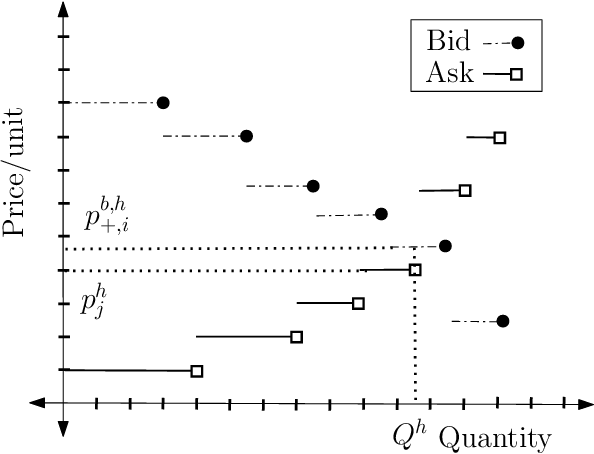}}
    \subfigure[]{\includegraphics[width=0.45\linewidth]{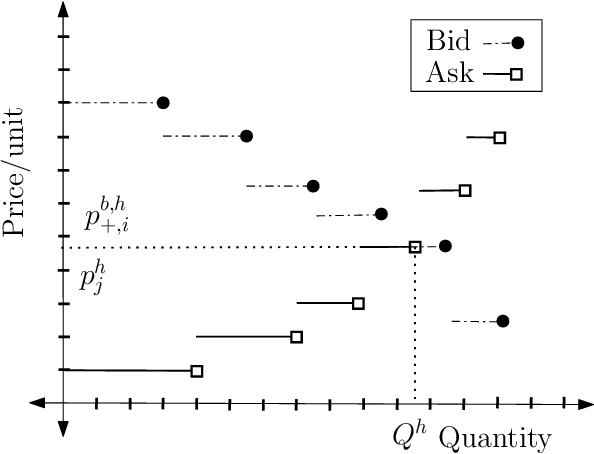}}
    \caption{Market clearing scenarios. (a) Supply available is greater than the demand. Bids and asks are cleared until demand is fully met. (b) Supply not enough to meet demand; Clearing happens until supply is exhausted. (c) and (d) Clearing happens until supply curve cross over the demand curve. In (d) the last cleared bid is placed at the same price as the last cleared ask. }
     \label{fig:clearingmechanism}
  \end{figure}

     The clearing process, as depicted in Figure \ref{fig:clearingmechanism}, is such that the highest buy bid priced  $p^{\buy,h}_{1,+}$ is \textit{matched} with the lowest ask (or sell bid) available at $\hat{p}^h_1$. Matching involves satisfying the demand $q^{\buy,h}_{1,+}$ with the supply $\hat{q}^h_1$. In case, the demand $q^{\buy,h}_{1,+}$ is fully met with the supply $\hat{q}^h_1$, the buy bid $(p^{\buy,h}_{1,+}, q^{\buy,h}_{1,+})$ is \textit{fully cleared} and any surplus supply $(\hat{q}^h_1 - q^{\buy,h}_{1,+})$ is matched with the next buy bid $(p^{\buy,h}_{2,+}, q^{\buy,h}_{2,+})$ if $\hat{p}^h_1 \leq p^{\buy,h}_{2,+}$. If the demand $q^{\buy,h}_{1,+}$ is more than the supply $\hat{q}^h_1$, the excess demand $(q^{\buy,h}_{1,+} - \hat{q}^h_1)$ is matched with the next ask (or sell bid)  $(\hat{p}^{h}_2, \hat{q}^{h}_2)$ if $p^{\buy,h}_{1,+} \geq \hat{p}^h_2$. In this way, costlier buy bids are cleared before the inexpensive ones and cheaper asks (or sell bids) are cleared before pricier ones until either (a) the demand is fully met (b) supply gets exhausted (c) the supply and demand curve cross each with a bid price becoming cheaper than some ask price (or sell bid price). Most often, the last cleared buy bid (ask/sell bid) is \textit{partially cleared} wherein the cleared quantity is less than the buy bid (ask/sell bid) quantity. Further, it is possible that multiple bids (asks/sell bids) could have been placed at the same price as the last cleared bid (ask/sell bid) and if the quantity component of such bids (asks/sell bids) are also same, all such bids (asks/sell bids) have same priority in the sense that the cleared quantity of all such bids (asks/sell bids) are same. Finally, bids (asks/sell bids) that are same in price but differ in quantity component are arranged in decreasing order of the bid (ask/sell bid) quantity.

\noindent \textbf{The $k$-double auction :}    
The average clearing price mechanism (ACPR) is a special case of $k$-Double auction with $k=0.5$ and its clearing price is given as $\lambda^h = \frac{\hat{p}^{h}_d +  p^{\buy,h}_{l,+} }{2}$. Additionally, the clearing price is the center point between the last cleared buy bid and ask (or sell bid) prices as shown in Figure \ref{fig:clearingmechanism}(a). 

\begin{lemma}
    The $k$-double auction satisfies the properties in Definition \ref{def:CM}. 
\end{lemma}
\begin{proof}
By construction the cleared quantities adhere to the properties defined in Definition \ref{def:CM}. The clearing price mentioned earlier in the section satisfies the condition $\lambda \in [\hat{p}^h_d,p^{\buy,h}_{l,+}]$. 
\end{proof}

\noindent \textbf{Merit order dispatch :}
The merit order  involves sequential clearing of bids and asks wherein the clearing is done by maximizing the area between supply and demand curve. More precisely, the dispatch mechanism can be modelled as an LP problem whose primal  solution 
$\bar{\boldsymbol{\alpha}}^h$ and  dual solution $\lambda^h$ 
gives the cleared quantities and clearing price of the all bids. The LP problem is formulated as  below :
\begin{equation}\label{eqn:linear_optimization}
    \begin{split}
    \min_{\bar{\boldsymbol{\alpha}}^{h}} \quad & {\boldsymbol{p}^{h}}^{\top}\cdot \bar{\boldsymbol{\alpha}}^{h}\\
    \textrm{s.t.} \quad & 	
    \bar{\boldsymbol{\alpha}}^{h}\le\boldsymbol{q}^{h}\\
        - &\bar{\boldsymbol{\alpha}}^{h} \le 0\\	&\Gamma^{h} \bar{\boldsymbol{\alpha}}^{h}=0 \;\;\; :\lambda^h
    \end{split}
    \end{equation}
     where 
    \begin{align*}
        \bar{\boldsymbol{\alpha}}^{h}&=(\alpha^{h}_1,\ldots,\alpha^{h}_d,\ldots,\alpha^h_{L^h+B^h_{-}}, \alpha^{i,h}_1,\ldots,\alpha^{\buy,h}_l,\ldots, \alpha^{j,h}_{B^h_{+}}),\\
        \boldsymbol{p}^h&=(p^h_1,\ldots,p^{h}_d,\ldots,p^h_{L^h+B^h_{-}}, -p^{i,h}_1,\ldots,-p^{\buy,h}_l,\ldots, -p^{j,h}_{B^h_{+}}),\\
       \boldsymbol{q}^h&=(q^h_1,\ldots,q^{h}_d,\ldots,q^h_{L^h+B^h_{-}}, q^{i,h}_1,\ldots,q^{\buy,h}_l,\ldots, q^{j,h}_{B^h_{+}}),\\ & \;\;\;\;\;\;\;\;\;\;\;\;\;\;\;\;\;\;\;\;\; i,j,\buy \in [N]\\
        \Gamma^{h}&= (-\vec{1}^{L^h+B^h_{-}},\vec{1}^{B^h_{+}})
    \end{align*}
    and $\lambda^h$ is the dual value corresponding to the equality constraint. Note that the indices of $\bar{\boldsymbol{\alpha}}^h$,$\boldsymbol{p}^h$ and $\boldsymbol{q}^h$ are taken from sets $\hat{\mathcal{L}}^h$ and $\mathcal{B}^h_{+}$ respectively.
    
    \begin{lemma}
        The LP problem of \eqref{eqn:linear_optimization} has a solution that satisfies the properties in Definition \ref{def:CM}. 
    \end{lemma}
    \begin{proof}
    The Lagrangian of the LP \eqref{eqn:linear_optimization}, is given as
    \begin{align*}    
        E(\bar{\boldsymbol{\alpha}}^h,\boldsymbol{\nu}^{h},\boldsymbol{\xi}^h,\lambda^h) &= {\boldsymbol{p}^{h}}^{\top}\cdot \bar{\boldsymbol{\alpha}}^{h} + {\boldsymbol{\nu}^h}^\top \cdot (\bar{\boldsymbol{\alpha}}^h-\boldsymbol{q}^h) - {\boldsymbol{\xi}^h}^\top \cdot \bar{\boldsymbol{\alpha}}^h \\ &+ \lambda^h \Gamma^h \bar{\boldsymbol{\alpha}}^h
    \end{align*}
    where the dual variables are
    \begin{align*}
        \begin{split}
             \boldsymbol{\nu}^{h}&=(\nu^{h}_1,\ldots,\nu^{h}_d,\ldots,\nu^h_{L^h+B^h_{-}}, \nu^{i,h}_1,\ldots,\nu^{\buy,h}_l,\ldots, \nu^{j,h}_{B^h_{+}}), \\
             \boldsymbol{\xi}^{h}&=(\xi^{h}_1,\ldots,\xi^{h}_d,\ldots,\xi^h_{L^h+B^h_{-}}, \xi^{i,h}_1,\ldots,\xi^{\buy,h}_l,\ldots, \xi^{j,h}_{B^h_{+}}), \\ 
             &\;\;\;\;\;\;\;\;\;\;\;\;\;\;\; i,j,\buy \in [N]\\
        \end{split}
    \end{align*}    
    The KKT conditions are
    \begin{align}\label{eqn:kkts}    
    \begin{split}
    \boldsymbol{p}^{h} + {\boldsymbol{\nu}^h}  - {\boldsymbol{\xi}^h} + \lambda^h \Gamma^h  &= 0\\
    \bar{\boldsymbol{\alpha}}^h - \boldsymbol{q}^h &\le 0\\
    -\bar{\boldsymbol{\alpha}}^h &\le 0\\
    \Gamma^h \bar{\boldsymbol{\alpha}}^h &=0\\
    \boldsymbol{\nu}^h &\ge 0\\
    \boldsymbol{\xi}^h &\ge 0\\
    \nu^h_e \cdot (\alpha^h_e-\hat{q}^h_e) &= 0, \;\; e = 1,\dots,L^h\\
    \nu^{\buy,h}_f \cdot (\alpha^{\buy,h}_{f}-q^{\buy,h}_{f,+}) &= 0, \;\; f = 1,\dots,B^h\\
    -\xi^h_e \cdot \alpha^h_e &= 0, \;\; e = 1,\dots,L^h\\
    -\xi^{\buy,h}_f \cdot \alpha^{\buy,h}_f &= 0, \;\; f = 1,\dots,B^h
    \end{split}
    \end{align}
    
    Now we have $\bar{\boldsymbol{\alpha}}$  as
    \begin{align}\label{eqn:admissible_criteria}
        \begin{split}
            \alpha^h_e &= \hat{q}^h_e,\text{ for } 1\le e<d_{\min}, \\
            \alpha^h_e &\le \hat{q}^h_e,\text{ for } d_{\min} \le e \le d_{\max}, \\
            \alpha^h_e &= 0,\text{ for } d_{\max} < e \le L^h+B^h_{-},  \\
            \alpha^{\buy,h}_f &= q^{\buy,h}_{f,+}, 1 \le f < l_{\min},\\
            \alpha^{\buy,h}_f &\le q^{\buy,h}_{f,+}, l_{\min} \le f \le l_{\max},\\
            \alpha^{\buy,h}_f &= 0, l_{\max} < f \le B^h_{+},\\               
        \end{split}
    \end{align}
    where $d_{\min} \le e \le d_{\max}$ denote the indices of the asks (sell bids) which are equal to the cleared ask (or sell bid) index  $d$ (refer Definition \ref{def:CM}) and similarly $l_{\min}$ and $l_{\max}$ are defined.
    The ask prices (and sell bid prices) $1 \le e < d_{\min}$ and buy bid prices $1\le f< l_{\min}$ are fully cleared. Hence from KKTs in \eqref{eqn:kkts} we have $\xi^h_e =0, \nu^h_e \ge 0$ and $\xi^{\buy,h}=0,\nu^{\buy,h} \ge 0$, which leads to $-\hat{p}^h_e + \nu^h_e - \lambda^h = 0$ and $-p^{\buy,h}_{f,+} + \nu^{\buy,h}_f - \lambda^h = 0$. Using these inequalities and equalities we have $p^{\buy,h}_{f,+} - \hat{p}^h_e = \nu^h_e + \nu^{\buy,h}_f > 0$ and this is true for possible values of $\nu^h_e > 0, \nu^{\buy,h}_f >0$.
    In the similar lines using \eqref{eqn:admissible_criteria} and \eqref{eqn:kkts}, we can show that for the cases $d_{\min} \le e \le d_{\max}, l_{\min} \le f \le l_{\max}$ and $d_{\max} < e \le L^h, l_{\max} < f \le B^h$, the inequalities are satisfied.
    
    Hence it can be concluded that the KKT conditions satisfy the properties defined in Definition \ref{def:CM}.
\end{proof}

\section{The Markov Game Structure}\label{sec:MGF}
    Consider a Markov game \cite{Zhang2023} with a finite horizon to model the PDA with $N$ prosumers. Specifically, let $\mathcal{G} = \langle N, S, A, C, P, H \rangle$ denote the Markov game with $N$ as the number of players (prosumers), $S$ as the state space, $A$ as the joint action space, $C = \{C^{\buy,h} \; | \; \buy \in [N],\; h \in [H]\}$ as the cost functions for the players, with $C^{\buy,h}$ denoting the cost function for the player $b$ at round $h$, $P: S \times A \rightarrow S$ as the transition probability and $H$ as the length of the horizon.     
    
    More concretely, the state space $S$ is defined as the  set of wholesale suppliers' asks $\mathcal{L}^h$, brokers' demand $\mathcal{Q}^h_{+}$ and brokers' supply $\mathcal{Q}^h_{-}$ and thus $S$ is $\{\mathcal{L}^h,\mathcal{Q}^h_{+},\mathcal{Q}^h_{-}\}$. Here, $\mathcal{Q}^h_{+}$ and $\mathcal{Q}^h_{-}$ are sets of brokers' demand $Q^{\buy,h}_{+}$ and supply $Q^{\buy,h}_{-}$ for all $\buy \in [N]$. Note that since we are working with the \emph{complete information setting}, the wholesale suppliers' asks, brokers' demand and brokers' supply are known to all the brokers;  hence, the transition of the states is deterministic. The joint action space $A = \times_{\buy \in [N]} A^{\buy}$ is a product space of all the prosumer's action space $A^{\buy}$, where $A^b = \cup_{h \in [H]} (\mathcal{B}^{\buy,h}_{+} \times \mathcal{B}^{\buy,h}_{-})$ is the product space of the all the buy bids $\mathcal{B}^{\buy,h}_{+}$ and sell bids $\mathcal{B}^{\buy,h}_{-}$ over $h \in [H]$. The cost function $C^{\buy,h} : S \times A^{\buy} \rightarrow \mathbb{R}$ for a player $\buy$ is given by
    \begin{align*}
        C^{\buy,h}(s^h, a^h ) = \begin{cases}
            \begin{aligned}[b]
                \bigg[ \sum\nolimits_{n\in [B^{\buy,h}_{+}]} \lambda^h \cdot \alpha^{\buy,h}_n \;\qquad \\ -  \sum\nolimits_{m\in [B^{\buy,h}_{-}]} \lambda^h \cdot \beta^{\buy,h}_m \bigg] 
            \end{aligned} &  0 \le h \le H\\
    \Upsilon \times Q^{\buy,h} &  h=H+1,
    \end{cases}
    \end{align*}
    where $a^h = (a^{\buy,h},a^{-\buy,h}) \in A$ denote the joint action, $a^{-\buy,h}$ denote the actions taken by players other than $\buy$, $\alpha^{\buy,h}_n$ are cleared quantities for player $b$'s buy bids and $\beta^{\buy,h}_m$ are the cleared sell bids. The constant $\Upsilon \ge 0$ is the \textit{balancing price} required to buy the quantity $Q^{\buy,H+1}_+$ outside of the PDA at time $H+1$. 
    
    A state $s^h$ at round $h$, when a joint action $a^h \in A$ is taken, transitions to a state $s^{h+1} = \{\mathcal{L}^{h+1},\mathcal{Q}^{h+1}_{+},\mathcal{Q}^{h+1}_{-}\}$ at round $h+1$. Here, $\mathcal{L}^{h+1}$ represents the uncleared asks at round $h+1$ and $\{\mathcal{Q}^{h+1}_{+}, \mathcal{Q}^{h+1}_{-}\}$ are the updated demand and supply respectively of the prosumers after accounting for the cleared quantities at round $h$. In addition, the sequence of transitions $s^1,a^1,s^2,a^2,\dots,s^H,a^H,s^{H+1}$, which starts at $s^1$ and ends at $s^{H+1}$, is denoted as a trajectory $\tau$ of the Markov game. Here, the trajectory $\tau$ induces a sequence of costs $C^{\buy,1},\dots, C^{\buy,H},C^{\buy,H+1}$ for each player $\buy \in [N]$. A player $\buy$ will choose an action $a^{\buy,h} \in A^b$ at each of the state $s^h$ she visits and the collection of actions taken at each round $h$ is denoted by a Markov policy $\pi^{\buy} = \{\pi^{\buy,h} : S \rightarrow  {A^\buy}\; |\; h \in [H] \}$. We further let $\pi = (\pi^\buy,\pi^{-\buy})$ as the joint policy that includes player $\buy$'s policy $\pi^{\buy}$ along with the policies of players except $\buy$, denoted as $\pi^{-\buy}$.  Moreover, we let $\Pi^{\buy}$ denote the policy space of player $\buy$ and $\Pi = \times_{\buy \in [N]} \Pi^{\buy}$ denote the joint policy space. Importantly, to capture the cost of acquisition, we define a value function $V^h_{\pi}: S \rightarrow \mathbb{R}$ of a joint policy $\pi \in \Pi$ at round $h$ in Equation~\eqref{eqn:value_fn_dfn}.
    \begin{equation}\label{eqn:value_fn_dfn}
        V_{\pi}^h(s) =  \sum\nolimits_{r=h}^{H+1} C^{\buy,r}(s^{r}, a^{\buy,r}, a^{-\buy,r}).   
     \end{equation}
     More specifically, the value function is given by 
     \begin{align*}
        V_{\pi}^h(s) = \sum\limits_{r=h}^{H} \lambda^r \cdot \left( \sum\limits_{n \in [B^{\buy,r}_{+}]} \alpha^{\buy,r}_n - \sum\limits_{m \in [B^{\buy,r}_{-}]} \beta^{\buy,r}_m\right) + \Upsilon \cdot Q^{\buy,H+1}_+. 
     \end{align*}
    For the game $\mathcal{G}$, we define MPNE \cite{abs-2011-00583} using the value function in Definition~\ref{defn:MPNE_defn}.    
     
    \begin{definition}\label{defn:MPNE_defn}
       Given the game $\mathcal{G}$ of $H$ horizon with $N$ players, a joint policy $\pi_* = (\pi^{\buy}_*, \pi^{-\buy}_*)$ is an MPNE if $\forall \buy \in [N]$, $\forall s \in S$, $\forall h \in [H]$ and $\forall \pi^\buy : S \rightarrow A^{\buy}$, we have 
       \begin{equation}
     V^h_{\pi_*^{\buy}, \pi_*^{-\buy}}(s) \le V^h_{\pi^{\buy}, \pi_*^{-\buy}}(s). 
     \label{eqn:Nashcondition}
    \end{equation}
    \end{definition}
    Having described the Markov game framework, we now proceed to develop MPNE solutions in the next section.  
    
\section{Equilibria of the Markov Game}\label{sec:nash}
    We begin by focusing on the case where the prosumers $\buy \in [N]$ are restricted to place at most one buy bid and one sell bid. In addition, first, we consider that the supply from the bulk sellers is enough to satisfy the demand, that is $Q^{\Sel,h} \ge Q^{\Buy,h}$. Later, we consider the inadequate supply case, where the equilibrium solution differs from the adequate supply case. Note that it is essential to consider the inadequate supply case since GenCos may not always be able to produce the supply needed to satisfy the expected demand. Furthermore, an incremental case of prosumers placing multiple buy and sell bids is considered. 

    \subsection{Adequate supply case}
        Let us define a few entities that will help to illustrate the proposed MPNE policy. First, denote $Q^{\Buy_{-\buy},h} = Q^{\Buy,h} -  Q^{\buy,h}$ as the demand of players excluding the player $\buy \in [N]$ at round $h$. Second, let $u_h$ be the lowest index of an ask in the sorted set $\mathcal{L}^h$ such that the supply up to the first $u_h$ asks is greater than the quantity $Q^{\Buy,h} - Q^{\Sel,h}_{-}$. That is, $u_h =\argmin\nolimits_{i} (Q^{\Buy,h} - Q^{\Sel,h}_{-} < \sum\nolimits_{m=1}^{i}q^h_m)$. 
        Similarly, let $v^{\buy}_{h}$ be an index of the sorted set $\mathcal{L}^h$ such that the supply up to first $v^{\buy}_{h}$ asks is greater than $Q^{-\buy,h}-Q^{\Sel,h}_{-}$. That is, 
        \begin{equation}\label{eqn:vhk}
            v^{\buy}_h = \argmin\nolimits_{j}\left(  Q^{\Buy_{-\buy},h} - Q^{\Sel,h}_{-} < \sum\nolimits_{m=1}^{j}\hat{q}^h_m \right) \; \forall \; \buy \in [N].    
        \end{equation}
        Finally, define an index  of $\mathcal{L}^h$ as $z_h = \max\{v^1_h, v^0_h \}$ and let $\phi^h$ be the player who bids at a price $p_{z_h}$, where $v^0_h$ is $u_h - (H-h)$ and $\phi^h$ is defined as,
        \begin{equation}\label{eqn:phi_defn}
            \phi^h =  \max\{1,\argmax\nolimits_{\buy}\{v^{\buy}_h \le v^0_h \}\}. 
        \end{equation}
    
    \paragraph{MPNE policy for adequate supply case:} The joint policy $\pi_*$ suggests that the player $\buy$ needs to check $Q^{\buy,h} > q_{u_h}$, that is, whether her selling quantity is greater than the ask quantity at the index $u_h$ (of $\mathcal{L}^h$). If yes, then the player places sell bid\footnote{Bid indices omitted since there is only one sell and one buy bid.} at price $p^{\buy,h}_{-,*} = p_{u_h}$ with quantity $q^{\buy,h}_{-,*} = Q^{\buy,h}_{-}$. However, if $Q^{\buy,h} \le q_{u_h}$ the prosumer places sell bid at price $p^{\buy,h}_{-} = p_{u_h} - \epsilon$, where $\epsilon >0$ is a small constant. Furthermore, the player has to check the condition $\buy = \phi^h$ and if the constraint is satisfied, then the player bids at buy bid price $p^{\buy,h}_{+,*}= p_{z_h}$ with quantity $q^{\buy,h}_{+,*} = Q^{\buy,h}_{+}$. However, when $\buy \neq \phi^h$, the player places a buy bid price $p^{\buy,h}_{+,*}=p_{\max}$ and quantity bid equal to her requirement. The MPNE policy  $\pi^{\buy,h}_*$ with buy and sell bids  $(p^{\buy,h}_{+,*}, \; q^{\buy,h}_{+,*}, \; p^{\buy,h}_{-,*}, \; q^{\buy,h}_{-,*} )$ is given in  Equation \eqref{eqn:MPNE1}. Note that the policy $\pi^{\buy,h}_*$ considers both sell bids and buy bids of the prosumer, unlike ~\cite{manvi2023nash}, where only buy bids are considered.  
    \begin{equation}\label{eqn:MPNE1}
    \pi^{\buy,h}_* =  \begin{cases}
            \begin{aligned}[b]
            \big(   p_{z_h}, \;   Q^{\buy,h}_{+} , p_{u_h}-\epsilon \mathds{1}_{\{Q^{\buy,h}_{-} \le q_{u_h}\}}, \; Q^{\buy,h}_{-} \big)
            \end{aligned}
             &  \textrm{if } \; \buy = \phi^h  \\
        \begin{aligned}[b]
            \big( p_{\max}, \;  Q^{\buy,h}_{+}, p_{u_h}-\epsilon \mathds{1}_{\{Q^{\buy,h}_{-} \le q_{u_h}\}}, \; Q^{\buy,h}_{-}  \big)
            \end{aligned}&   \textrm{o.w.}  
        \end{cases}    
    \end{equation}

    Having explained the candidate policy $\pi_*$, the value function of the joint policy $\pi^{\buy,h}_*$ at round $h$ for player $\buy \in [N]$ at state $s^h \in S$ is given in Equation~\eqref{eqn:value_fn} using the properties of the uniform clearing mechanism defined in Definition \ref{def:CM}. 
    
    \begin{equation}\label{eqn:value_fn}
      V^h_{\pi^{\buy}_*,\pi^{-\buy}_*}(s) =
    \begin{cases}\!
      \begin{aligned}[b]
        \bigg[ p_{z_h} \cdot  Q^h - p_{z_h} \cdot \sum\nolimits_{i \in [N] \setminus \phi^h} q^{i,h}_j \\ -  p_{z_h} \cdot \beta^{\buy,h}_*  
            + \sum\nolimits_{k=h+1}^{H} p_{z_{k}} \cdot Q^{k} \\ -  \sum\nolimits_{k=h+1}^{H} p_{z_{k}} \cdot \beta^{\buy,h}_* \bigg],
      \end{aligned} & \textrm{if } \buy = \phi^h \\
      p_{z_h} \cdot ( Q^{\buy,h} - \beta^{\buy,h}_*), & \textrm{o.w.} 
    \end{cases} 
    \end{equation}

    The cleared quantities and market clearing price for the proposed MPNE policy \eqref{eqn:MPNE1} is given by following Lemma.
    
\begin{lemma}\label{lem:clearingstatistics}
   If at round $h$, the  supply from bulk sellers is adequate to satisfy the requirement of all players, that is, $Q^{\Buy,h} \leq Q^{\Sel,h}$ and if all the players adopt the policy $\pi_*^{\buy,h}$ in Equation \eqref{eqn:MPNE1}, then we have
\begin{enumerate}
    \item  The total market cleared quantity at round $h$ is \[Q^h_* = \min\left(\sum\nolimits_{j=1}^{z_h} \hat{q}^h_j, \sum\nolimits_{\buy \in [N]} Q^{\buy,h}\right)\].  
    \item The buy bid placed by the player $\buy \neq \phi_h$, at round $h$ gets fully cleared. That is, $\alpha^{\buy,h}_* =  Q^{\buy,h},\;\; \forall \buy \neq \phi_h$. 
    \item The buy bid placed by the player $\buy = \phi_h$ at round $h$, gets cleared as,         $\alpha^{\buy,h}_* =  \left(Q^h - \sum\nolimits_{i \in [N] \setminus \phi_h} q^{i,h} \right)$. 
     \item The sell bid for player $\buy$ may or may not clear at round $h$. Because, the bid price of the bids may not intersect with sell bid. However, before the end of $H$ rounds, the sell bid (of all players $\buy \in [N]$) is guaranteed to be cleared at a price $\lambda^r_*$ for $r \in [h\;;H]$ with $\beta^{\buy,r}_*$ cleared quantity.
\end{enumerate}
\end{lemma}
\begin{proof}
First note that  policy $\pi_*^{\buy,h}$ in Equation \eqref{eqn:MPNE1} has just two price bids with the highest bid price at $p_{\max} \geq p^h_{L^h+B^h_{-}}$. This ensures that at least bid is cleared and hence $Q^h > 0$.  Now in case of adequate supply from bulk sellers (that is  $Q^{\Buy,h} \leq Q^{\Sel,h}$), the player $\phi^h$ will bid at price $p_{z_h}$. By construction,  $p_{z_h}$ is also the point where the supply and demand curve intersect and hence the $\lambda^h$ is $p_{z_h}$. It is now easy to see that, the total market cleared quantity is given by, 
\[Q^h = \min\left(\sum\nolimits_{j=1}^{z_h} \hat{q}^h_j, \sum\nolimits_{\buy \in [N^h]} Q^{\buy,h}\right).\]
As the bids placed at the higher price $p_{\max}$ gets cleared first and since the available supply is enough to cater to the  outstanding demand requirement at round $h$, bids gets cleared exactly as stated in the Lemma.

Finally, the sell bids are placed such that the they are fully cleared but at a highest possible clearing price to make sure the the player gets better price for selling in the auction. With this in mind, the sell bids are either fully cleared at price $p_{u_h}$ or $p_{u_h}-\epsilon$ depending on their sell bid price. Hence, by design the sell bids are guaranteed to be cleared within $H$ rounds. 
\end{proof}

    \paragraph{Equilibrium analysis: }
    We demonstrate that the candidate policy in Equation~\eqref{eqn:MPNE1} is indeed MPNE policy in the space of all deterministic policies. 
    More specifically, we show that, for all players $\buy \in [N]$, for all state $s^h \in S$, for all round $h \in [H]$ and for any deterministic policy $\pi^\buy$, Equation \eqref{eqn:Nashcondition} is satisfied.
    To this end, we consider the value function for all possible deviations where the sell bid deviations are tabulated in Table \ref{tbl:deviations}.  Furthermore, the buy bid deviations are exactly the same as in Table \ref{tbl:deviations} (except for the notation from ${p}^{\buy,h}_{-}$ to ${p}^{\buy,h}_{+}$). Note that, by assumption, prosumers cannot buy more than they require and sell more than they have; hence, only five sell and buy bid deviations are possible. Finally, the combined deviations $\pi^{\buy} \in \Pi^{\buy}$ is the cartesian product of the sell bid deviations and the buy bid deviations\footnote{In total, $25 (5\times 5)$ deviations are possible.}. In the sequel, we provide a preliminary result in the Lemma \ref{lem:prelim}. The first part of Lemma \ref{lem:prelim} provides a property of the clearing price $\lambda^h$ for a uniform clearing rule, which satisfies  Definition \ref{def:CM}. The second part of Lemma \ref{lem:prelim} provides a condition on the cost of balancing outside the horizon of the PDA. 

    \begingroup
    \renewcommand{\arraystretch}{1.4}
    \begin{table}[ht]
    \centering
    \caption{Possible Sell Bid Deviations}
    \label{tbl:deviations}
    \begin{tabular}{|cc|}
    \hline
    \multicolumn{2}{|c|}{Equal Priced Deviation}                               \\ \hline
    \multicolumn{2}{|c|}{$p^{\buy,h}_{-} = p^{\buy,h}_{-,*}$,  $q^{\buy,h}_{-} < q^{\buy,h}_{-,*}$}                              \\ \hline
    \multicolumn{1}{|c|}{Higher Priced Deviations}                                               & Lower Priced Deviations                                              \\ \hline
    \multicolumn{1}{|c|}{$p^{\buy,h}_{-} > p^{\buy,h}_{-,*}$,  $q^{\buy,h}_{-} < q^{\buy,h}_{-,*}$} & $p^{\buy,h}_{-} <p^{\buy,h}_{-,*}$,  $q^{\buy,h}_{-} < q^{\buy,h}_{-,*}$ \\ \hline
    \multicolumn{1}{|c|}{$p^{\buy,h}_{-} > p^{\buy,h}_{-,*}$, $q^{\buy,h}_{-} = q^{\buy,h}_{-,*}$}  & $p^{\buy,h}_{-} < p^{\buy,h}_{-,*}$, $q^{\buy,h}_{-} = q^{\buy,h}_{-,*}$ \\ \hline
    \end{tabular}
    \end{table}
    \endgroup
    
\begin{lemma}
\label{lem:prelim}
Given the supply curve from bulk sellers across rounds  $h\in [H]$ is constant and the clearing mechanism satisfies the properties defined in Definition \ref{def:CM}, we have the following.
\begin{enumerate}
    \item \label{st:MCP}  The clearing price at $h$ satisfy $\lambda^h \ge \lambda^{h+1}$ for $h \in [H-1]$.
    \item \label{st:bal_price} The condition on the price\footnote{The price is also called balancing price.} to buy outside the auction at $H+1$ denoted by $\Upsilon$ is given by 
$\Upsilon > \gamma \cdot p_{\max}$, where $\gamma >1$.
\end{enumerate}
\begin{proof}

  \textbf{Proof of statement \ref{st:MCP}}:
  Recollect that the clearing price $\lambda^h$ is unique once a clearing mechanism adhering to properties in Definition  \ref{def:CM} is chosen. The clearing price belong to an interval $[\hat{p}^h_d;p^{\buy,h}_{+}]$, where $\hat{p}^h_d$ is the last cleared ask (or sell bid) at round $h$. Since the ask (or sell bid) is either partially or fully cleared, the lower limit of the interval would stay same (in case of partially cleared) or increase (in case when fully cleared). 

 \textbf{ Proof of statement \ref{st:bal_price}:}
  Here, for any deviated policy $\pi^{\buy}$ by the player $\buy$ when all other players play Nash policy $\pi^{-\buy}_*$, the value function has to satisfy \eqref{eq:value_fn_bal} for $\pi_*$ to be an MPNE.
    \begin{equation}\label{eq:value_fn_bal}
    \begin{split}
     V^h_{\pi_*^{\buy}, \pi_*^{-\buy}}(s) &\le V^h_{\pi^{\buy}, \pi_*^{-\buy}}(s) \\
     V^h_{\pi_*^{\buy}, \pi_*^{-\buy}}(s) - V^h_{\pi^{\buy}, \pi_*^{-\buy}}(s) &\le 0\\
     \sum\nolimits_{r=h}^H \left(\alpha^{\buy,r}_* \lambda^r_* -\alpha^{\buy,r} \lambda^r\right) - \Upsilon Q^{b,H+1} &\le 0\\
    \end{split}
    \end{equation}
    Here, the only high priced deviations or the combination of high priced and low priced deviations can lead to $Q^{\buy,H+1} =0$. Hence the value of $\Upsilon$ does not matter in this case. In addition, in case of  higher priced deviations alone we can show that $ \sum\nolimits_{r=h}^H \left(\alpha^{\buy,r}_* \lambda^r_* -\alpha^{\buy,r} \lambda^r\right) \le 0$. 
    
    Now with the combination of low priced and equal priced deviations the remaining quantity is $Q^{\buy,H+1} > 0$. In this case, the \textit{worst case} condition on $\Upsilon$ is such that $\Upsilon > \frac{Q_{\max} \cdot p_{\max}}{q_{\min}} + p_{\max}$ which can be written as $\Upsilon > \gamma \cdot p_{\max}$, where $Q_{\max}$ is the maximum quantity that is needed by any player and $q_{\min}$ is the resolution (minimum sold quantity) of the auction market. Here, for all practical cases the the range of $Q_{\max}$ is lesser than the range of quantity $q_{\min}$, hence $\gamma$ is a reasonable finite number.
\end{proof}
\end{lemma}

    We now move to the Theorem \ref{thm:main}, which shows that the value function of the deviated policies is greater than the candidate policy's value function.

 \begin{theorem}
\label{thm:main}
Given the conditions in Lemma \ref{lem:prelim} hold, we have the following.
\begin{enumerate}
    \item \label{st:sell_bid_deviation} The sell bid deviations $\pi^{\buy} \in \Pi^{\buy}$ of player $\buy$ with any buy bids $p^{\buy,h}_{+} \in [0;p_{\max}]$ and $q^{\buy,h}_{+} \in [0;Q^{\buy,h}_{+}]$ satisfy \eqref{eqn:value_deviation}. 
    \item \label{st:buy_bid_deviation} The buy bid deviations $\pi^{\buy} \in \Pi^{\buy}$ of player $\buy$ with any sell bids $p^{\buy,h}_{-} \in [0;p_{\max}]$ and $q^{\buy,h}_{-} \in [0;Q^{\buy,h}_{-}]$ satisfy \eqref{eqn:value_deviation}.
\end{enumerate}
\begin{align}\label{eqn:value_deviation}
        V^h_{\pi_*^{\buy}, \pi_*^{-\buy}}(s) \le V^h_{\pi^{\buy}, \pi_*^{-\buy}}(s)
\end{align}

\end{theorem}
\begin{proof}
\textbf{Proof of statement \ref{st:sell_bid_deviation}: }
Recall that the value function of any deterministic policy $\pi$ in PDA is given by 
\begin{align*}
    V_{\pi}^h(s) = \sum\limits_{r=h}^{H} \lambda^r \cdot \left( \sum\limits_{n \in [B^{\buy,r}_{+}]} \alpha^{\buy,r}_n - \sum\limits_{m \in [B^{\buy,r}_{-}]} \beta^{\buy,r}_m\right) + \Upsilon \cdot Q^{\buy,H+1}.
 \end{align*}
whereas the value of the candidate MPNE is given in Equation \eqref{eqn:MPNE1}. The value function for the case of one sell bid and one buy bid reduces to
\begin{align*}
    V_{\pi}^h(s) = \sum\limits_{r=h}^{H} \lambda^r \cdot \left(  \alpha^{\buy,r} -  \beta^{\buy,r}\right) + \Upsilon \cdot Q^{\buy,H+1}.
 \end{align*}

 We note the use of remark \ref{rem:bid_sell_no_match} here, in case of low priced deviations, the only way a player $\buy$ can get lower value function is when her buy bid is matched with her own sell bid, where the sell bid is the only element in $\hat{\mathcal{L}}^h$. In this situation, the player $\buy$ can manipulate the market by placing sell bid  close to zero  price and at some buy bid $p^{\buy,h}$. This at round $h< H$ with condition $p_{z_{h+1}} <p_{z_{h}}(1+0.5 \frac{Q^{\buy,h}_{-}}{Q^{\buy,h}_{-}+Q^{\buy,h}_{+}})$ can lead to lower cost (value function to player $\buy$). However, by remark \ref{rem:bid_sell_no_match} this is not allowed to happen. Furthermore, the assumption that $[N^h] \subseteq [N]$ ensures that the no new player can join the auction later. Moreover, by design there will be no new players at $H+1$ to buy the remaining sell bids. Hence if the sell bids are not cleared before $H+1$, then  they will never be cleared. 

With above remarks, we show that at $h=H$ the value function\footnote{We suppress the state in the notation of Value function.} $V^H_{\pi}=\lambda^H \cdot (\alpha^{\buy,H}-\beta^{\buy,H}) + \Upsilon \cdot Q^{\buy,H+1}_+$ of all sell bid deviations at round $h=H$ is greater than or equal to the candidate policy's value function $V^h_{\pi_*} = p_{z_H} \cdot (Q^{\buy,H}_{+} - Q^{\buy,H}_{-})$ (here $\beta^{\buy,H}_* = Q^{\buy,H}_{-}$) . To this end, for equal priced deviations the sell bid quantity cleared to the player $\buy$  is $\beta^{\buy,H} < \beta^{\buy,H}_*$. Hence the value function is $V^H_{\pi^{\buy},\pi^{-\buy}} = p_{z_H} \cdot (\alpha^{\buy,H}-\beta^{\buy,H}) + \Upsilon \cdot Q^{\buy,H+1}_+$, where $\beta^{\buy,H} < \beta^{\buy,H}_*$ implies $V^H_{\pi_*} \le V^H_{\pi}$. 
Next for high priced deviations, there are two possibilities: one where player deviates above the  price $p_{u_H}-\epsilon$ and other is when player bids above the price $p_{u_H}$. In both cases, the sell bid will not clear since by design the sell bids loose their priority to the asks from $\mathcal{L}^H$ and hence, the value function is $V^H_{\pi} = p_{z_H} \cdot \alpha^{\buy,H} + \Upsilon \cdot Q^{\buy,H+1}_+ > V^H_{\pi_*}$. Finally, the low priced deviations recommend to bidding below the price $p_{u_H}-\epsilon$ or $p_{u_H}$. Since the player is underbidding, the player's sell bid is fully cleared at same price $P_{z_H}$, however as the buy bids can vary we might have $\alpha^{\buy,h} \le \alpha^{\buy,h}_*$. Hence the player's value function is $V^H_{\pi} = p_{z_H} \cdot (\alpha^{\buy,H} - \beta^{\buy,H}) + \Upsilon \cdot Q^{\buy,H+1} \ge  V^H_{\pi_*}$. 

Now let the value function at $h=O+1$ satisfies $V^{O+1}_{\pi_*} \le V^{O+1}_{\pi}$ for all $s \in S$. Aim is to show function function satisfies $V^{O}_{\pi_*} \le V^{O}_{\pi}$ at $h=O$. To this end, consider the expression
\begin{align*}
    V^O_{\pi} = \lambda^O \cdot (\alpha^{\buy,O} - \beta^{\buy,O}) + V^{O+1}_{\pi}  
\end{align*}
Now from the arguments made at the $h=H$, all deviations will yield the immediate cost as $\lambda^O \cdot (\alpha^{\buy,O} - \beta^{\buy,O}) \ge \lambda^O_* \cdot (\alpha^{\buy,O}_* - \beta^{\buy,O}_*)$. Hence with assumption of $V^{O+1}_{\pi} \ge V^{O+1}_{\pi_*}$ we have $V^{O}_{\pi} \ge V^{O}_{\pi_*}$, which implies by induction that $V^{h}_{\pi} \ge V^{h}_{\pi_*} \; \forall h \in [H],s \in S, \pi^{\buy} \in \Pi^{\buy}$. 

\textbf{Proof statement \ref{st:buy_bid_deviation}: }
Similar to statement \ref{st:sell_bid_deviation} of Theorem \ref{thm:main}, we first show the buy bid deviations yield higher cost compared to the candidate MPNE policy at $h= H$. To this end, for equal priced deviations the cleared buy bid quantity for player $\buy$ is $\alpha^{\buy,H} \le \alpha^{\buy,H}_*$. Hence the value function \footnote{Again, the state is suppressed in value function notation} for this deviation (with any sell bid deviation) is $V^h_{\pi} = \lambda^H \cdot (\alpha^{\buy,H} - \beta^{\buy,H}) + \Upsilon \cdot Q^{\buy,H+1}_+ \ge V^H_{\pi_*}$. Next, for higher priced deviations (this deviation is not allowed when player recommended to bid at $p_{\max}$) the clearing price $\lambda^H \ge \lambda^H_*$  would increase there by increasing the value function. That is, $V^H_{\pi} = \lambda^H \cdot (\alpha^{\buy,H} - \beta^{\buy,H}) \ge V^H_{\pi_*}$. Finally, for lower priced deviations, the cleared quantity might lower with lower clearing price. However, it might lead to buying non zero quantity outside auction at $H+1$, which is very expensive by Statement  \ref{st:bal_price} of Lemma \ref{lem:prelim}. Hence the value function is $V^H_{\pi} \ge V^H_{\pi_*}$

Now assume at round $h=O+1$ the value function satisfies $V^{O+1}_{\pi} \ge V^{O+1}_{\pi_*}$ for all $s \in S$. Similar to Statement  \ref{st:sell_bid_deviation} of this Theorem, we aim to show $V^O_{\pi} \ge V^{O}_{\pi_*}$ at $h=O$. However, unlike statement \ref{st:sell_bid_deviation} of this Theorem , we need to argue differently to show the requirement. To this end, for the equal price deviation at $h=O$ the cleared quantity decreases, that is $\alpha^{\buy,O} \le \alpha^{\buy,O}_*$. This implies that the cost at round $O$ decreases compared to candidate policy as $\lambda^O \cdot (\alpha^{\buy,O}-\beta^{\buy,O}) \le \lambda^O_* \cdot (\alpha^{\buy,O}_*-\beta^{\buy,O}_*)$. However,  using Statement\footnote{The Statement \ref{st:MCP} of Lemma \ref{lem:prelim} says $\lambda^{r} \ge \lambda^{O}$, where $r >O$.} \ref{st:MCP} of Lemma \ref{lem:prelim}  and letting $\delta = \alpha^{\buy,O}_* - \alpha^{\buy,O}$, we have value function at $O$  as $V^O_{\pi} = \lambda^O \cdot (\alpha^{\buy,O}-\beta^{\buy,O}) + \lambda^r \cdot \delta + V^{O+1}_{\pi} \ge V^{O}_{\pi_*}$, where $r > O$. Next, the higher priced deviations (when the player is not bidding $p_{\max}$) at round $O$ will have a cost $\lambda^O \cdot (\alpha^{\buy,O}-\beta^{\buy,O}) \ge \lambda^O_* \cdot (\alpha^{\buy,O}_*-\beta^{\buy,O}_*)$. Hence by using $V^{O+1}_{\pi} \ge V^{O+1}_{\pi_*}$ and the immediate cost for higher priced deviation we have  $V^{O}_{\pi} \ge V^{O}_{\pi_*}$.  Finally, the lower priced deviations will have a cost $\lambda^O \cdot (\alpha^{\buy,O}-\beta^{\buy,O}) \le \lambda^O_* \cdot (\alpha^{\buy,O}_*-\beta^{\buy,O}_*)$ because of lower cleared quantity $\alpha^{\buy,h} \le \alpha^{\buy,h}_*$ and  lower clearing price $\lambda^h \le  \lambda^h_*$. Here, using condition on balancing price $\Upsilon$ and clearing price $\lambda^r, \; r > O$ from \ref{lem:prelim} we have that buying the quantities later will be either costlier or remain same. That is, for this deviation the value function is $V^{O}_{\pi} = \lambda^O \cdot (\alpha^{\buy,O} - \beta^{\buy,O}) + \delta \cdot \lambda^r + V^{O+1}_{\pi} \ge V^{O}_{\pi_*}$.

\end{proof}
    
\subsection{Inadequate supply}\label{sec:not_enough_supp}
    We now consider the inadequate supply,  
    (that is, $Q^{\Sel,h} < Q^{\Buy,h}$). To provide the MPNE for this case, we modify the definitions of $u_h$, $\phi^h$, $v^k_h$ and $z_h$ provided in the adequate supply case. First, we assign $u_h = L^h$ as the value. Second, $\phi^h$ is given as 
    \begin{equation}\label{eqn:phi_defn_nosup}
        \phi^h = 
            \argmin\nolimits_j\{ Q^{\Sel,h} \le \sum\nolimits_{\buy = 1}^{j} Q^{\buy,h}\} 
    \end{equation}
    Third, when $\phi^h = N$, we set index $v^N_{h}$ as 
    \begin{equation*}
    v^N_h = \argmin_j \left( Q^{\Buy_{-N},h} - Q^{\Sel,h}_{-} \le \sum_{m=1}^j q^h_m\right). 
    \end{equation*}
    Finally, $z_h$ when $\phi^h = N$ is defined as $z_h = \max\{v^{\phi}_h, v^0_h\}$. However, if $\phi^h < N$, then $p_{z_h} = p_{\max}$.
    
    \paragraph{MPNE policy for inadequate supply case:} The buy bids mentioned in Equation \eqref{eqn:MPNE1} now use modified definitions, whereas the sell bids change to bid at a price $p_{-}^{\buy,h}=\Upsilon - \epsilon$ and bid quantity $q^{\buy,h}_{-} = Q^{\buy,h}_{-}$. We provide the value function of the inadequate supply in Equation \eqref{eqn:value_fn_inadqt}.

\begin{equation}\label{eqn:value_fn_inadqt}
      V^h_{\pi^{\buy}_*,\pi^{-\buy}_*}(s) =
    \begin{cases}\!
      \begin{aligned}[b]
        p_{z_h} \cdot  Q^h \\ - p_{z_h} \cdot \sum\nolimits_{i \in [N] \setminus \phi^h} q^{i,h}_j \\ - p_{z_h} \cdot \beta^{\buy,h}_*)  \\ 
        + \sum\nolimits_{k=h+1}^{H} p_{z_{k}} \cdot Q^{k} \\ - \sum\nolimits_{k=h+1}^{H} p_{z_{k}} \cdot \beta^{\buy,h}_* \\ + \Upsilon \cdot Q^{\buy,H+
1}_+ 
      \end{aligned} & \textrm{if } \buy = \phi^h = N \\
      p_{z_h} \cdot ( Q^{\buy,h} - \beta^{\buy,h}_*), & \buy < \phi^h \\

\begin{aligned}
      p_{z_h} \cdot (\alpha^{\buy,h}_* - \beta^{\buy,h}_*) \\ + \Upsilon \cdot Q^{\buy,H+
1}_+ 
\end{aligned} &\buy = \phi^h, \phi^h < N \\
\Upsilon \cdot Q^{\buy,H+
1}_+  & \buy > \phi^h \\
    \end{cases} 
    \end{equation}

Note that for player $\buy > \phi^h$ equation $Q^{\buy,H+1}_+ = Q^{\buy,h}$ holds.

\begin{lemma}
\label{lem:inadequate}
    The modified candidate policy is an MPNE.   
\end{lemma}
\begin{proof}
First, it can be seen that for player $\buy > \phi^h$ any deviations will not impact the value function hence all deviations are indifferent. 

Now, for players $\buy < \phi^h$, with help of value function in Equation \eqref{eqn:value_fn_inadqt}, the similar arguments made in Theorem \ref{thm:main} can be extended here to conclude that any deviation is expensive. Furthermore, for player $\buy = \phi^h$ and $\phi^h =  N$, the arguments are same as in the Theorem \ref{thm:main}.  Finally, when player $\buy = \phi^h$ and $\phi^h < N$, the player deviating below the price $p_{\max}$ would lead to decrease in the priority and hence the player has to buy more quantity outside the auction at higher price $\Upsilon$. Hence, by Lemme \ref{lem:prelim}, we have that the cost of procurement is expensive.
\end{proof}

\subsection{ Multiple bid Equilibrium policy}
We now consider the case where the prosumers are allowed to place multiple buy and sell bids. We show that the the policy in \eqref{eqn:MPNE1} is an equilibrium policy even when multiple buy and sell bids are allowed. To this end, we provide the following Lemma. 
\begin{lemma}
    The MPNE in \eqref{eqn:MPNE1} which has single buy bid and single sell bid is still an MPNE when multiple buy bids and multiple sell bids are allowed.
    \begin{proof}
        The multi-bid policy where the quantity is divided among multiple bids leads to the quantity deviation of $q^{\buy,h} < q^{\buy,h}_*$ (both sell and buy bid) as shown in \ref{tbl:deviations}. Similar to  the proof of theorem \ref{thm:main}, we argue that the deviation will lead to higher cost. Observe that without division of quantity, the (true) multiple bids are not possible, hence the value function of all possible multiple bids is higher than the value function of the proposed MPNE.  
    \end{proof}
\end{lemma}

\section{Experimental Evaluation}
\label{sec:exp}

    In the previous sections, we presented analytical equilibrium solutions for PDAs using a Markov game framework. In doing so, we had a complete information setup wherein the players (prosumers) have knowledge of the supply curve and the demand information of other players at each round of the PDA. In this section, we use the equilibrium solutions \eqref{eqn:MPNE1} obtained in  the previous section to propose a bidding strategy \nash\ for the incomplete information case where a buyer has neither the knowledge of the demand requirement of other players nor complete information about the supply curve.  We start by providing the algorithm for \nash, followed by numerical experiments to showcase the efficacy of the proposed strategy. The numerical experiments are conducted on the wholesale market module of the Power Trading Agent Competition (PowerTAC)~\cite{Ketter2020}. 
  
    \noindent\textbf{PowerTAC: }PowerTAC is an efficient and close-to real-world smart grid simulator that models all the crucial elements of a typical smart grid system, including GenCo and energy brokers acting as prosumers. During the simulation (or game), which typically lasts for $60$ simulation days, an energy broker has to compete against several other brokers. To test our bidding strategy, we focus on PowerTAC's wholesale market PDAs, where the energy broker plays a crucial role in buying/selling energy. These PDAs are day-ahead auctions in which the broker can purchase energy $24$ hours ahead of the delivery timeslot by participating in a total of $24$ auctions at an interval of every hour. PowerTAC PDA employs ACPR and uniform pricing rules for clearing and payments. To bid in PowerTAC PDA, a broker must submit the bid price and quantity (decided based on demand forecasts). For determining the bid price, a broker may use the information available from the server, which includes the market-clearing price, its own cleared quantity, net cleared quantity and orderbook information. Orderbook includes an anonymized list of uncleared asks and bids, while the knowledge about the other brokers' cleared bids/asks is kept hidden. Failing to acquire the required quantity from the wholesale market, a broker has to purchase the remaining quantity at balancing market prices, which are typically high and serve as a penalty for a broker for causing imbalance. The simulation also includes a buyer called MISO that procures energy for a region that contains retailer users not serviced by the main brokers of the PowerTAC setup. The MISO buyer's energy requirement is almost ten times the PowerTAC retail market's energy requirements and substantially affects the clearing prices. The MISO buyer purchases all of its estimated demand in the first round of the PDA  and any excess procurement is sold in the subsequent rounds. The MISO always places a market order to buy or sell energy in any round of the PDA. The GenCos are the primary sellers in the market that follow a quadratic cost function to decide the ask prices. Refer to PowerTAC specifications~\cite{Ketter2020} for more information.

    \paragraph{\nash\ Algorithm: }
    We propose an algorithm \nash\ for the incomplete information setting of PowerTAC inspired by the equilibrium analysis on the complete information setting. As our proposed \nash\ bidding algorithm uses some design ideas from VV21, we first describe the design of VV21. It models the cost supply curve of the GenCos from the uncleared ask information data available from the simulator. The idea is to locate the price corresponding to the broker's bid quantity (requires demand forecast of broker's and market's demands) on the supply curve, treat that price as the upper bound on the limit-prices, and place multiple bids below that price. The reason for placing multiple bids below the chosen upper bound is that it aims to procure the majority of the quantity from the asks of the MISO buyer and other prosumers in the market and treats GenCo as the supplier in the last resort. The supply curve of the GenCo also has an element of randomization between rounds of the PDA; hence, placing multiple bids in a range-bound manner helps the agent procure energy at a lower price. More details of the VV21 strategy can be found in \cite{ijcai2022p23}.
    
    On the other hand, \nash\ decides the price on the supply curve by using the forecasts of the total demand and the broker's demand to determine the indices $u_h$ and $v^{\buy}_h$ on the uncleared asks list (number of uncleared asks to satisfy the demand). Using these indices, it estimates the best possible limit-price, which follows the analytical solution presented in Section~\ref{sec:nash}. After deciding the price, it places multiple bids similar to VV21 to work around the supply curve's randomness. Unlike VV21, \nash\ aims to purchase most of the quantity from GenCo.

    \begin{algorithm}[tb]
    \small
    \caption{\nash}
    \label{alg:nash_policy} 
        \begin{algorithmic}[1] 
        \STATE totalDmd[] $\leftarrow$ netDmdPredict(currentTime)
        \STATE $Q^{\buy,h}[] \leftarrow$ indvDmdPredictor[currentTime]
        \FOR{hour in $[1,\dots, 23]$}
        \STATE futureTime $\leftarrow$ currentTime + hour
        \STATE unclearedAsks[] $ \leftarrow$ Auction(currentTime-1,futureTime)
        
        \IF{unclearedAsks is not empty}
            \STATE $Q^{\Sel,h} \leftarrow$ sum(unclearedAsks.q)
            \STATE $Q^{\Buy,h} \leftarrow$ totalDmd[futureTime]
                \IF{$Q^{\Sel,h} \ge Q^{\Buy,h}$}
                    \STATE $p_{u_h} \leftarrow \min p_r $ s.t   $Q^{\Buy,h} \le \sum_{i=1}^{r} \text{unclearedAsks}.q_i$
                    \STATE $Q^{\Buy_{-\buy},h}  \leftarrow$ $Q^{\Buy,h} - Q^{\buy,h}$[futureTime]
                    \STATE $p_{v^{\buy}_h} \leftarrow \min p_r $ s.t  $Q^{\Buy_{-\buy},h} \le \sum_{i=1}^{r} \text{unclearedAsks}.q_i$
                \ELSE 
                    \STATE  $p_{u_h} \leftarrow p_{L^h}$, $p_{v^{\buy}_h} \leftarrow P_{L^h}$
                \ENDIF
            \STATE $v^0_h \leftarrow \max\{1,u_h - \text{hour} + 1\}$
            \STATE estBidPrice = $\max\{p_{v^0_h}, p_{v^{\buy}_h}\}$
        \ELSE
            \STATE clearedPrices[] $\leftarrow$ Auction($t$,$t$+hour) $\forall t <$ currentTime
            \STATE estBidPrice = $\max$$\{$cleredPrices$\}$
        \ENDIF
        \IF{(currentTime is far from futureTime)}
            \STATE minP = $\alpha_f$ * estBidPrice; maxP = $\beta_f$ * estBidPrice
        \ELSE 
            \STATE minP = $\alpha_c$ * estBidPrice; maxP = $\beta_c$ * estBidPrice
        \ENDIF
        \STATE Sample $P$ prices, $p^{\buy,h}_i \sim$ U[minP,maxP]
        \STATE Distribute $Q^{\buy,h}$ uniformly across $P$ prices, $q^{\buy,h}_i \leftarrow Q^{\buy,h}/P$
        \STATE bidList $\leftarrow (p^{\buy,h}_i,q^{\buy,h}_i) \; \forall \; i \in \{1, 2, ... , P\}$
        \STATE Auction(currentTime,futureTime) $\leftarrow$ bidList
        \ENDFOR
        \end{algorithmic}
    \end{algorithm}

    As shown in the algorithm, \nash\ takes the current bidding timeslot as input, uses the $24$th hour (the first opportunity) of each auction to observe the uncleared asks, and places a list of bids for each of the next $23$ hours. For each of these future hours (futureTime), the algorithm queries for the list of uncleared asks from past auction data. These asks are a list of price and quantity tuples $(p_i, q_i)$ sorted in increasing order of prices. Then, based on the list of unclear asks, the algorithm estimates the bid price (estBidPrice) by utilising the knowledge of the market and its own demand forecasts. The bid price estimation approach in lines 4 to 21 of algorithm \ref{alg:nash_policy} is inspired by the equilibrium solution presented in Equation \eqref{eqn:MPNE1}. As the buyer has $24$ opportunities to procure the required demand, thus can afford to take risks during the initial rounds and play conservatively in the last few rounds, as shown in lines $23$ to $27$. The hyperparameters $\alpha_f$, $\beta_f$, $\alpha_c$ and $\beta_c$ can be decided based on the risk level of the buyer. After sampling $P$ prices uniformly, the buyer's required quantity is uniformly divided into $P$ bids and submitted for clearing. 

    \begin{figure*}
      \centering
      \includegraphics[width=0.9\textwidth]{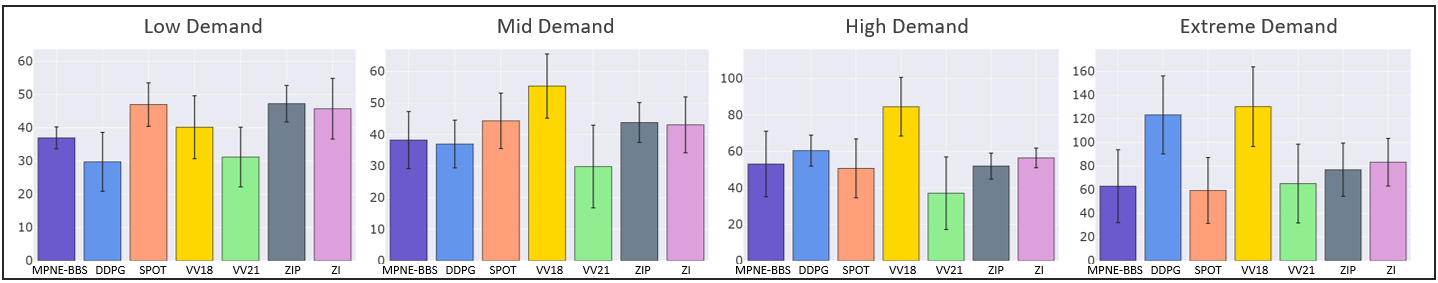}
      \caption{Wholesale Cost Comparison in 7-Player Games (with MISO Buyer)}
      \label{fig:7p_w_miso}
    \end{figure*}
    
    \begin{figure*}
      \centering
      \includegraphics[width=0.9\textwidth]{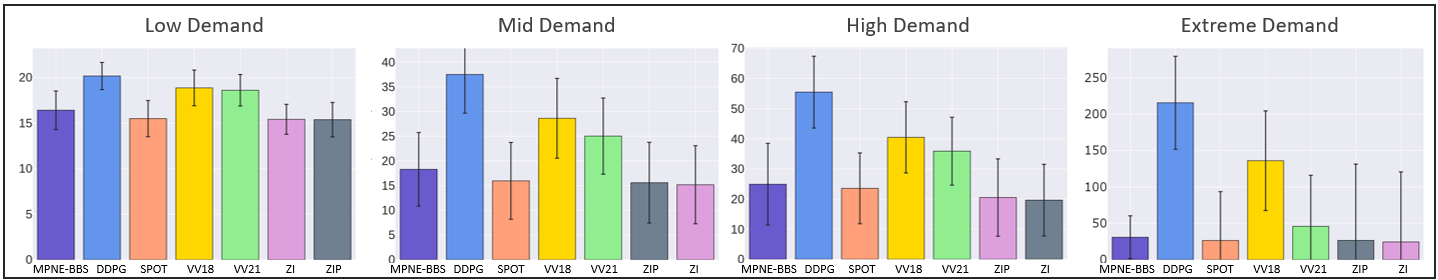}
      \caption{Wholesale Cost Comparison in 7-Player Games (without MISO Buyer)}
      \label{fig:7p_wo_miso}
    \end{figure*}
    
    \noindent\textbf{Benchmark: }Below, we briefly describe the state-of-the-art as well as baseline bidding strategies of PowerTAC PDAs that are used in the experiments to compare the performance of \nash.

    \begin{itemize}
        \item \textbf{VV21}: VidyutVanika21 (VV21)~\cite{ijcai2022p23} aims to model the cost curve of the GenCos to identify the lowest ask and use it to place suitable bids in the auctions. It handles the randomness of the lowest ask from the cost curve by heuristically placing multiple bids around the lowest ask region. VV21 is the best-performing bidding strategy for PowerTAC PDAs~\cite{ijcai2022p23}.
        \item \textbf{DDPGBBS}: Deep-Deterministic-Policy-Gradient-based Bidding Strategy (DDPGBBS)~\cite{sanjay22} is based on the Bayesian Nash equilibrium for two-unit double auctions and exercises a scale-based bidding strategy. It uses a DDPG-based RL strategy to learn the equilibrium scale factors to decide the bids.   
        \item \textbf{SPOT}: SPOT~\cite{spot18} uses a Monte Carlo Tree Search (MCTS) based bidding strategy, which is integrated with a REPTree-based limit price predictor. On top of it, SPOT uses a few heuristics techniques to determine the optimal bid prices and place multiple bids in an auction.        
        \item \textbf{VV18}: VidyutVanika18 (VV18)~\cite{susobhan20} employs an MDP-based bidding strategy, which is solved using dynamic programming with the help of limit-price predictor to decide bids. Additionally, it heuristically spreads the quantity across $24$ instances based on the predicted limit prices.          
        \item \textbf{ZI}: The Zero Intelligence (ZI) strategy is a randomized approach that ignores any information about the current state of the market or the value of the item being auctioned and places a single bid per auction instance by sampling a bid price from a heuristically decided uniform distribution.
        \item \textbf{ZIP}: The Zero Intelligence Plus (ZIP)~\cite{zip} computes the bid price by multiplying a unit limit price with a scalar variable $m$ denoting the profit it aims to achieve. Small increments adjust the price for each trade with the help of a $\delta$ by comparing the submitted bid price and the clearing price. 
        
    \end{itemize}

    \noindent\textbf{Experimental Setup: }To test the efficacy of \nash\ against the above-listed bidding strategies, we perform two sets of experiments on the PowerTAC platform, one that includes the MISO buyer and the other without it. In these experiments, we compare the unit purchase costs of the brokers in the wholesale market, a strategy having a lower purchase cost is preferable. In Set-1, we play all-player games that include the above six bidding strategies and \nash, along with MISO buyer in the market. We ask all the players except MISO to procure a fixed demand requirement for each timeslot; the demand requirement is the same for all the brokers for a fair comparison. Set-1 is further divided into four configurations depending on the demand requirements of the brokers, namely, low-demand, mid-demand, high-demand and extreme-demand. Similarly, in Set-2, we remove the MISO buyer from the market and repeat the same experiments for all four demand levels. We play $10$ games for each configuration in each set (Figures ~\ref{fig:7p_w_miso} and  \ref{fig:7p_wo_miso}). Additionally, we play two-player games between \nash\ and all the other strategies; that is, we play $10$ games between \nash\ and VV21, $10$ games between \nash\ and ZI, and so on. We perform these experiments for both the sets and all four demand levels as mentioned above (Tables~\ref{tab:2p_w_miso} and \ref{tab:2p_wo_miso}).

    \noindent\textbf{Results: }As shown in Figure ~\ref{fig:7p_w_miso}, \nash\ is one of the best-performing bidding strategies in terms of wholesale cost as it achieved close to the best or second-best cost among the seven brokers across variable demand levels. Specifically, its wholesale cost is close to the best wholesale cost in the market for extreme demand level (only $6\%$ away from the best) and close to second-best for high and mid demand levels ($4.7\%$ and $3.4\%$ difference, respectively), with only VV21 performing better. The performance is more prominent in Set-2; as shown in Figure~\ref{fig:7p_wo_miso}, it achieves a wholesale cost very close to the best wholesale cost in the market across all demand levels and consistently outperforms DDPG and VV21. 

    \begin{table}[t]
        \small
        \centering
        \caption{Performance in 2-Player Games (with MISO Buyer)}
        \label{tab:2p_w_miso}
        \begin{tabular}{| p{1.4cm} | p{1.2cm} | p{1.2cm} | p{1.2cm} | p{1.2cm}|}
        \hline
        \centering\textbf{Opponent} & \centering\textbf{Low} & \centering\textbf{Mid} & \centering\textbf{High} & \textbf{Extreme} \\
        \hline
        \centering\textbf{VV21} & \centering{\textbf{1.01}} & \centering{\textbf{0.98}} & \centering{\textbf{1.03}} & \hspace{0.4cm}\textbf{1.18}\\
        \hline
        \centering\textbf{DDPG} & \centering{0.73}  & \centering{0.85} & \centering{\textbf{1.12}} & \hspace{0.4cm}\textbf{1.64} \\
        \hline
        \centering\textbf{SPOT} & \centering{\textbf{1.45}}  & \centering{\textbf{1.09}}  & \centering{0.87} & \hspace{0.4cm}0.77 \\
        \hline
        \centering\textbf{VV18} & \centering{\textbf{1.32}}  & \centering{\textbf{1.36}}  & \centering{\textbf{1.49}}  & \hspace{0.4cm}\textbf{1.76} \\
        \hline
        \end{tabular}
    \end{table}

   The results of the 2-Player experiments in Tables~\ref{tab:2p_w_miso} and \ref{tab:2p_wo_miso} show the wholesale cost of the opponent strategy relative to \nash, a value more than $1$ would indicate \nash\ is the superior bidding strategy among the two. Particularly, it matches VV21, which is the best strategy in the literature and achieves similar wholesale cost as VV21 in all demand levels, with and without MISO buyer. It even outperforms VV21 for the extreme demand level where VV21's wholesale cost is $1.18$ times higher than \nash's cost. Overall, the results show that it achieves superior performance against each opponent and for almost all demand levels. This set of experiments aims to validate the efficacy of \nash\ against several different state-of-the-art strategies having different bidding patterns. Furthermore, we performed the same experiments for 3-Player and 5-Player games. On an average, the relative costs of best and worst opponent strategies were $1.13$ and $1.87$, respectively in 5-player without MISO games; $1.0$ and $2.34$, respectively in 5-player with MISO games. Similarly, for 3-player games, these numbers are $1.27$ and $1.47$, respectively for best and worst opponent strategies without MISO; $0.98$ and $1.19$, respectively with MISO. Thus, the simulation results show that the \nash\ achieves significant performance improvements against the best state-of-the-art bidding strategy across various number of and players in a game for various market and demand scenarios. 

   \begin{table}[t]
        \small
        \centering
        \caption{Performance in 2-Player Games (without MISO Buyer)}
        \label{tab:2p_wo_miso}
        \begin{tabular}{| p{1.4cm} | p{1.2cm} | p{1.2cm} | p{1.2cm} | p{1.2cm}|}
        \hline
        \centering\textbf{Opponent} & \centering\textbf{Low} & \centering\textbf{Mid} & \centering\textbf{High} & \textbf{Extreme} \\
        \hline
        \centering\textbf{VV21} & \centering{\textbf{0.99}}  & \centering{\textbf{0.99}} & \centering{\textbf{1.0}} & \hspace{0.4cm}\textbf{0.99} \\
        \hline
        \centering\textbf{DDPG} & \centering{\textbf{1.08}}  & \centering{\textbf{1.07}} & \centering{\textbf{1.02}} & \hspace{0.4cm}\textbf{1.08} \\
        \hline
        \centering\textbf{SPOT} & \centering{0.91}  & \centering{0.82}  & \centering{\textbf{1.08}} & \hspace{0.4cm}\textbf{1.09} \\
        \hline
        \centering\textbf{VV18} & \centering{\textbf{1.04}}  & \centering{\textbf{1.15}}  & \centering{\textbf{1.16}}  & \hspace{0.4cm}\textbf{1.13} \\
        \hline
        \end{tabular}
    \end{table}

\section{Conclusion}\label{sec:conclusion}
    In this paper, we proposed equilibrium strategies for prosumers involved in a PDA to buy and sell commodities. We modelled the PDA as a Markov game with prosumers as players and derived equilibrium solutions for the complete information setting when the players are aware of the supply curve and the outstanding demand requirement of other players at every round of the PDA. Specifically, we derived MPNE solutions for the setting when the players compete to procure required commodities. Thereafter, the proposed MPNE solutions were used to design a bidding algorithm called \nash\ for a more practical setup and its efficacy was demonstrated using the PowerTAC simulator test-bed against several state-of-the-art algorithms. In future work, we hope to extend the analysis of equilibrium solutions to the incomplete information setting by modelling the PDA as a partially observable stochastic (Markov) game and compare its effectiveness against the algorithm proposed in this work. 


\bibliographystyle{named}
\bibliography{ijcai24}

\end{document}